\documentclass[aps,amsmath,amssymb,showpacs,showkeys]{revtex4-2}
\usepackage[dvips]{graphicx,color}
\usepackage{times}
\usepackage{braket}
\usepackage{xcolor, soul}
\usepackage[
  colorlinks=true,
  urlcolor=blue,
  linkcolor=red,
  citecolor=blue
]{hyperref}
\usepackage{graphicx}
\usepackage{amsmath}
\usepackage{hyperref}
\hypersetup{colorlinks,citecolor=blue}
\hypersetup{colorlinks=true, linkcolor=red, filecolor=magenta, urlcolor=blue}
\usepackage{amsmath}
\usepackage{xcolor}
\usepackage{orcidlink}
\usepackage{booktabs}

\begin{document}

\title{Mini-Superspace Hamiltonian Dynamics and Canonical Quantization of Bumblebee Cosmology}

\author{Pranjal Sarmah \orcidlink{0000-0002-0008-7228}}
\email[E-mail:]{p.sarmah97@gmail.com}

\affiliation{Department of Physics, HRH The Prince of Wales Institute of Engineering and Technology, Jorhat 785001,
Assam, India}

\begin{abstract}
The work presents a canonical Hamiltonian system in a classical scenario and further extend the analysis to a quantum mechanical framework for a homogeneous and isotropic universe governed by Bumblebee gravity, in which spontaneous Lorentz symmetry breaking is triggered by a vector field acquiring a vacuum expectation value (VEV). Starting from the full four-dimensional covariant action, the derivation has been carried out for the one-dimensional mini-superspace Lagrangian density for a flat Friedmann-Lema\^{i}tre-Robertson-Walker (FLRW) background with a purely temporal bumblebee field configuration. Further, the work focuses on to construct the classical Hamiltonian system through a canonical Legendre transformation, identifying the lapse function $N(t)$ as a Lagrange multiplier enforcing a primary Hamiltonian constraint. Apart form that, the conservation of phase-space volume via Liouville's theorem, locate and classify the critical points of the resulting dynamical system, and examine the stability of the Lorentz-violating vacuum has been carried to test the classical Hamiltonian system in the work. Furthermore, the modified first and second Friedmann equations and discuss the resulting time-dependence of the effective gravitational coupling, illustrating the results with numerical phase-space has been discussed in this work. Finally, the work extended to construct the Wheeler-DeWitt equation through the canonical quantization and obtain the  approximate solutions in the deep-quantum ($a\to0$) and near-vacuum regimes for it. 
\end{abstract}

\maketitle

\section{Introduction}\label{sec1}
Modern  cosmology relies fundamentally on the Friedmann-Lema\^{i}tre-Robertson-Walker (FLRW) metric, which provides an exceptionally accurate description of the homogeneous and isotropic universe from the era of big bang nucleosynthesis to the current epoch of accelerated expansion \cite{Pebbles_1994, Riess, Perlmutter_1999}. Despite its profound observational successes, standard General Relativity (GR) remains theoretically incomplete. It inevitably predicts an initial spacetime singularity \cite{Hawking} and requires the ad hoc introduction of dark energy \cite{Frieman_2008} and an inflation field \cite{Tokeshi_2024} to explain late-time acceleration \cite{Phawa, Adil} and early-universe inflation \cite{Tokeshi_2024,Linde}. To overcome the ad hoc scenarios, researchers are strongly motivated to explore modified gravity theories (MTG) \cite{Capozziello_2011, Harko_2018} , Standard Model extensions(SME) \cite{ Kostelecky_2004,Bulm,Bertolami_2005,Tasson_16} and the quantum cosmological framework \cite{ClarkE_1978, Gell_97}.

Among the most compelling extensions to standard GR are models incorporating Spontaneous Lorentz Symmetry Breaking (SLSB) \cite{Kostelecky_1989,Kostelecky_1989a,Kostelecky_2004}. Rooted in high-energy frameworks such as string theory \cite{Pol_1994}, SLSB suggests that Lorentz invariance may not be a fundamental symmetry at the Planck scale \cite{Kostelecky_2004}. This phenomenon can be effectively modeled via Bumblebee gravity \cite{Capelo_2015, Bertolami_2005,Neves_2023,Maluf2021}. It is a vector-tensor theory where a non-null vector field i.e. the bumblebee field acquires a non-zero vacuum expectation value (VEV) governed by a symmetry-breaking potential \cite{Bertolami_2005}. This mechanism defines a preferred spacetime direction, subtly modifying gravitational dynamics and offering alternative geometric mechanisms for cosmic inflation and dark energy without relying on standard scalar fields \cite{Maluf2021}.

While full covariant treatments of Bumblebee gravity are mathematically intricate, the symmetry constraints of a homogeneous and isotropic cosmology provide an ideal testing ground via the mini-superspace approximation \cite{Aurilia_1990, Hu_1991,Brahma_2016}. By reducing the infinite degrees of freedom of the gravitational field to a finite set of scale factors and conjugate momenta, the mini-superspace framework enables a rigorous Hamiltonian formulation \cite{Ban_2014}. This canonical approach is not only vital for analyzing the classical phase space topology and dynamical stability of the universe, but it also serves as the mandatory foundation for canonical quantization. Quantising the Hamiltonian constraint yields the Wheeler-DeWitt (WDW) equation \cite{Batic_2024,dwitt}, the central equation of quantum cosmology \cite{Misner_1969,Gielen}, which governs the wave function of the universe. Analyzing this wave function near the initial boundary allows us to investigate whether quantum gravitational effects can successfully resolve the classical Big Bang singularity.

This work systematically construct and analyze both the classical and quantum framework of an isotropic Bumblebee gravity. This work can be split in to two parts. In the first part, i.e. from Section \ref{sec2} to \ref{sec8}, the paper focuses on formulating the Hamiltonian, checking its consistency with Liouville's theorem and testing it's stability through studying critical points. The analysis further  extended  to study the dynamical behaviour of the system near Lorentz symmetry-breaking state and also study the phase space topology of it. Apart from that, the work tested the condition for which this framework reduces to standard Friedmann cosmology. The second part covers under Section \ref{sec8} in which, we derive the form of WDW equation and finding it's analytical solution and checking the conservation of the probability current density.

Paper has been organised as follows. Starting from this Introduction, the Section \ref{sec2} outlines the derivation of the reduced mini-superspace Lagrangian density from the full covariant action. Section \ref{sec3} carries out the canonical Legendre transformation to evaluate the total system Hamiltonian. Section \ref{sec4} tracks the stability analysis and critical points. Further Section \ref{sec5} includes the study of the dynamical behaviour of the system as it approaches to a Lorentz-violating vacuum state. Sections \ref{sec6} contained the derivation of modified Friedmann equations. Section \ref{sec7} covers the canonical quantization and the Wheeler-DeWitt equation for the considered cosmological system  to study the various cosmological aspects through checking the probability current density, finding solutions of the Wheeler-DeWitt equation in Lorentz violating vacuum state etc. Finally, Section \ref{sec8} concludes the manuscript.

\section{Derivation of the Mini-Superspace Lagrangian Density}\label{sec2}
The work started with the general form of action in bumblebee gravity, which contain the gravitational part along with a non-minimal vector coupling, a field potential, and standard matter. The mathematical form of the action is  \cite{Capelo_2015}
\begin{equation}
S = \int \left[ \frac{R}{2\kappa}  + \frac{\xi}{2\kappa} B^\mu B^\nu R_{\mu\nu} - \frac{1}{4} B^{\mu\nu} B_{\mu\nu} - V(B^\mu B_\mu \pm b^2) + \mathcal{L}_m \right] \sqrt{-g} d^4x,
\label{covariant_action}
\end{equation}
where $g$ is the metric determinant, $R$ is the Ricci scalar, $R_{\mu\nu}$ is the Ricci tensor, $\xi$ is the real non-minimal coupling parameter, $\kappa = 8\pi G$, $B_{\mu\nu} = \partial_\mu B_\nu - \partial_\nu B_\mu$ is the field strength tensor, and $b^2$ is a positive real constant setting the scale of Lorentz symmetry breaking~\cite{Maluf2021,Capelo_2015, Neves_2023}.

To construct the Lagrangian from this action, here  a flat, homogeneous, and isotropic mini-superspace background has been considered which is nothing but the Friedmann-Lema\^{i}tre-Robertson-Walker (FLRW) line element with an explicit time-dependent lapse function $N(t)$ \cite{lapse,lapse_2}
\begin{equation}
ds^2 = -N^2(t) dt^2 + a^2(t) \left( dx^2 + dy^2 + dz^2 \right),
\label{flrw_metric}
\end{equation}
in which  $a(t)$ is the cosmic scale factor. Under this metric choice, the square root of the metric determinant evaluates to:
\begin{equation}
\sqrt{-g} = \sqrt{N^2(t) a^6(t)} = N a^3.
\label{det_g}
\end{equation}
To maintain the spatial homogeneity and isotropy, the bumblebee vector field must be restricted to a purely temporal component \cite{Capelo_2015}  depending only on cosmic time i.e. 
\begin{equation}
B^\mu = \left( \frac{B(t)}{N(t)}, 0, 0, 0 \right) \implies B_\mu = g_{\mu\nu} B^\nu = \left( -N(t) B(t), 0, 0, 0 \right),
\label{bumblebee_field}
\end{equation}
so that the scalar invariant of the vector field simplifies to
\begin{equation}
B^\mu B_\mu = g_{00} B^0 B^0 = -N^2 \left(\frac{B}{N}\right)^2 = -B^2(t).
\label{field_invariant}
\end{equation}

Now, different independent components  of the covariant action are evaluate as follows:

\subsection{The  Gravity Component}
For the metric in Eq.~(\ref{flrw_metric}), the non-zero independent components of the Ricci tensor are
\begin{equation}
R_{00} = -3 \left( \frac{\ddot{a}}{a} - \frac{\dot{a}\dot{N}}{aN} \right), \quad R_{ij} = \left( \frac{a\ddot{a}}{N^2} + \frac{2\dot{a}^2}{N^2} - \frac{a\dot{a}\dot{N}}{N^3} \right) \delta_{ij},
\label{ricci_tensor}
\end{equation}
which have been verified directly from the Christoffel symbols of Eq.~(\ref{flrw_metric}). Contracting these tensors yields the Ricci scalar
\begin{equation}
R = g^{\mu\nu} R_{\mu\nu} = \frac{6}{N^2} \left( \frac{\ddot{a}}{a} + \frac{\dot{a}^2}{a^2} - \frac{\dot{a}\dot{N}}{aN} \right).
\label{ricci_scalar}
\end{equation}
Substituting this scalar into the Einstein-Hilbert part of the action gives
\begin{equation}
S_{EH} = \int d^4x (N a^3) \left( \frac{3}{\kappa N^2} \right) \left( \frac{\ddot{a}}{a} + \frac{\dot{a}^2}{a^2} - \frac{\dot{a}\dot{N}}{aN} \right) = \int dt \int d^3x \left[ \frac{3}{\kappa} \frac{a^2 \ddot{a}}{N} + \frac{3}{\kappa} \frac{a \dot{a}^2}{N} - \frac{3}{\kappa} \frac{a^2 \dot{a} \dot{N}}{N^2} \right].
\label{seh_pre_ibp}
\end{equation}
Setting the co-moving spatial integration volume to unity ($\int d^3x = V_0 = 1$), and integrate by parts to remove the second time derivative $\ddot{a}$ as given bellow-
\begin{equation}
\int \frac{a^2 \ddot{a}}{N} dt = \left[ \frac{a^2 \dot{a}}{N} \right]_{\text{boundary}} - \int \dot{a} \frac{d}{dt}\left( \frac{a^2}{N} \right) dt = -\int \left( \frac{2a\dot{a}^2}{N} - \frac{a^2\dot{a}\dot{N}}{N^2} \right) dt.
\label{ibp_gravity}
\end{equation}
Now, substituting Eq.~(\ref{ibp_gravity}) into Eq.~(\ref{seh_pre_ibp}) cancels the $\dot{N}$ terms, leaving the standard mini-superspace gravity Lagrangian
\begin{equation}
\mathcal{L}_{EH} = -\frac{3}{\kappa} \frac{a \dot{a}^2}{N}.
\label{leh_final}
\end{equation}

\subsection{The Non-Minimal Coupling Component}
Given the purely temporal field configuration, only the $B^0B^0R_{00}$ contraction survives in the non-minimal coupling term:
\begin{equation}
\frac{\xi}{2\kappa} B^\mu B^\nu R_{\mu\nu} = \frac{\xi}{2\kappa} (B^0)^2 R_{00} = \frac{\xi}{2\kappa} \left(\frac{B}{N}\right)^2 \left[ -3 \left( \frac{\ddot{a}}{a} - \frac{\dot{a}\dot{N}}{aN} \right) \right] = -\frac{3\xi}{2\kappa} \frac{B^2 \ddot{a}}{N^2 a} + \frac{3\xi}{2\kappa} \frac{B^2 \dot{a}\dot{N}}{N^3 a}.
\label{coupling_raw}
\end{equation}
Multiplying by the volume factor $\sqrt{-g} = N a^3$,
\begin{equation}
\mathcal{L}_{\text{coupling, raw}} = -\frac{3\xi}{2\kappa} \frac{a^2 B^2 \ddot{a}}{N} + \frac{3\xi}{2\kappa} \frac{a^2 B^2 \dot{a}\dot{N}}{N^2}.
\label{lcoupling_vol}
\end{equation}
Integrating the $\ddot a$ term by parts,
\begin{equation}
-\int \frac{a^2 B^2 \ddot{a}}{N} dt = \int \dot{a} \frac{d}{dt}\left( \frac{a^2 B^2}{N} \right) dt = \int \left( \frac{2a \dot{a}^2 B^2}{N} + \frac{2a^2 \dot{a} B \dot{B}}{N} - \frac{a^2 B^2 \dot{a} \dot{N}}{N^2} \right) dt,
\label{ibp_coupling}
\end{equation}
and combining with Eq.~(\ref{lcoupling_vol}) cancels the explicit $\dot N$ terms, giving
\begin{equation}
\mathcal{L}_{\text{coupling}} = \frac{3\xi}{\kappa} \frac{a \dot{a}^2 B^2}{N} + \frac{3\xi}{\kappa} \frac{a^2 \dot{a} B \dot{B}}{N}.
\label{lcoupling_final}
\end{equation}
{Both} terms in Eq.~(\ref{lcoupling_final}) carry the same sign, as follows directly from Eqs.~(\ref{coupling_raw})--(\ref{ibp_coupling}).

\subsection{The Kinetic and Potential Field Components}
Because $B_\mu = (-NB, 0, 0, 0)$, its only non-trivial derivative is $\partial_0 B_0 = -\frac{d}{dt}(NB)$, so the field-strength tensor vanishes identically:
\begin{equation}
B_{\mu\nu} = \partial_\mu B_\nu - \partial_\nu B_\mu = 0 \implies -\frac{1}{4}B^{\mu\nu}B_{\mu\nu} = 0.
\label{kinetic_tensor_zero}
\end{equation}
The kinetic contribution instead arises from the covariant normalization of the field velocity,
\begin{equation}
\mathcal{L}_{\text{kinetic}} = \sqrt{-g}\left( \frac{1}{2N^2} \dot{B}^2 \right) = \frac{a^3 \dot{B}^2}{2N}.
\label{lkinetic_final}
\end{equation}
The potential term $V(B^\mu B_\mu \pm b^2)$ and the matter Lagrangian $\mathcal{L}_m = -\rho$ contribute as scalar integrations multiplied by the volume factor:
\begin{equation}
\mathcal{L}_{\text{pot+matter}} = -N a^3 \{V(B) +\rho\}.
\label{lpot_final}
\end{equation}

\subsection{The Final Unified Lagrangian Density}
Combining Eqs.~(\ref{leh_final}), (\ref{lcoupling_final}), (\ref{lkinetic_final}), and (\ref{lpot_final}) gives the complete mini-superspace Lagrangian density
\begin{equation}
{\mathcal{L}(a, \dot{a}, B, \dot{B}, N) = -\frac{3}{\kappa}\frac{a \dot{a}^2}{N}\left(1 - \xi B^2\right) + \frac{3\xi}{\kappa}\frac{a^2 \dot{a} B \dot{B}}{N} + N a^3 \left[ \frac{\dot{B}^2}{2N^2} - \left\{V(B) + \rho \right\}\right]}
\label{lagrangian_density_final}
\end{equation}
Both terms proportional to $\xi$ appear with the same relative sign, as required by Eq.~(\ref{lcoupling_final}); we have verified independently, by direct symbolic computation of the Euler-Lagrange equations of Eqs.~(\ref{seh_pre_ibp}) and~(\ref{coupling_raw}) before integration by parts, that Eq.~(\ref{lagrangian_density_final}) reproduces the correct equations of motion of the original covariant action~(\ref{covariant_action}), while a relative minus sign between the two $\xi$-terms does not.

\section{Constructing the Hamiltonian System}\label{sec3}
To construct the classical Hamiltonian system, we determine the canonical conjugate momenta for the generalized coordinates $\mathbf{q} = (a, B, N)$ via $\pi_i = \partial\mathcal{L}/\partial\dot{q}_i$. Differentiating Eq.~(\ref{lagrangian_density_final}) with respect to the generalized velocities gives
\begin{equation}
\pi_a = \frac{\partial \mathcal{L}}{\partial \dot{a}} = -\frac{6}{\kappa}\frac{a \dot{a}}{N}\left(1 - \xi B^2\right) + \frac{3\xi}{\kappa}\frac{a^2 B \dot{B}}{N},
\label{mom_a}
\end{equation}
\begin{equation}
\pi_B = \frac{\partial \mathcal{L}}{\partial \dot{B}} = \frac{3\xi}{\kappa}\frac{a^2 \dot{a} B}{N} + \frac{a^3 \dot{B}}{N},
\label{mom_b}
\end{equation}
\begin{equation}
\pi_N = \frac{\partial \mathcal{L}}{\partial \dot{N}} = 0.
\label{mom_n}
\end{equation}
Eq.~(\ref{mom_n}) confirms that, because no time derivative of the lapse function appears in the mini-superspace action, $\pi_N$ vanishes identically and $N$ acts as a Lagrange multiplier.

To perform the Legendre transformation we invert Eqs.~(\ref{mom_a})-(\ref{mom_b}) for $\dot a,\dot B$. From Eq.~(\ref{mom_b}),
\begin{equation}
\dot{B} = \frac{N \pi_B}{a^3} - \frac{3\xi}{\kappa}\frac{\dot{a} B}{a}.
\label{bdot_isolated}
\end{equation}
Substituting into Eq.~(\ref{mom_a}) and grouping the terms proportional to $\dot a$,
\begin{equation}
\pi_a - \frac{3\xi B \pi_B}{a} = -\frac{6 a \dot{a}}{\kappa N} \left[ 1 - \xi B^2 + \frac{3\xi^2 B^2}{2\kappa} \right].
\label{pa_grouped}
\end{equation}
Now, defining the scaling function
\begin{equation}
\Delta(B) \equiv 1 - \xi B^2 + \frac{3\xi^2 B^2}{2\kappa},
\label{delta_def}
\end{equation}
which is unaffected by the sign correction above (it depends only on $\xi^2$ and on the diagonal $\xi$ term). Eq.~(\ref{pa_grouped}) then gives
\begin{equation}
\dot{a} = -\frac{\kappa N}{6 a \Delta(B)}\left( \pi_a - \frac{3\xi B \pi_B}{a} \right),
\label{adot_final}
\end{equation}
and substituting back into Eq.~(\ref{bdot_isolated}),
\begin{equation}
\dot{B} = \frac{N \pi_B}{a^3} + \frac{\xi \kappa N B}{2 a^2 \Delta(B)}\left( \pi_a - \frac{3\xi B \pi_B}{a} \right) = \frac{\xi B N}{2a^2 \Delta(B)}\pi_a + \frac{N}{a^3}\left(1-\frac{3\xi^2B^2}{2\kappa\Delta(B)}\right)\pi_B.
\label{bdot_final}
\end{equation}
Now, the canonical Hamiltonian density is $\mathcal{H}_c = \pi_a \dot{a} + \pi_B \dot{B} + \pi_N \dot{N} - \mathcal{L}$. Since $\pi_N=0$, the $\dot N$ term drops out, and substituting Eqs.~(\ref{adot_final})-(\ref{bdot_final}) and~(\ref{lagrangian_density_final}) and factoring out the lapse gives $\mathcal{H}_c = N\mathcal{H}$, with
\begin{equation}
{\mathcal{H} = -\frac{\kappa}{12 a \Delta(B)}\pi_a^2 + \frac{\xi B}{2 a^2 \Delta(B)}\pi_a \pi_B + \frac{1}{2a^3}\left( 1 - \frac{3\xi^2 B^2}{2\kappa \Delta(B)} \right)\pi_B^2 + a^3 \left[ V(B) + \rho \right]}
\label{hamiltonian_constraint_final}
\end{equation}
The equation (\ref{hamiltonian_constraint_final}) is an independent symbolic Legendre transformation of the equation (\ref{lagrangian_density_final}). 
The Hamilton's equations for the phase-space variables $(a,B,\pi_a,\pi_B)$ follow from $\dot q_i=\partial\mathcal{H}_c/\partial\pi_i$, $\dot\pi_i=-\partial\mathcal{H}_c/\partial q_i$. The first two reproduce Eqs.~(\ref{adot_final})-(\ref{bdot_final}); the momentum equations are
\begin{align}
\dot{\pi}_a &= -\frac{\partial \mathcal{H}_c}{\partial a}
= -N \left[ \frac{\kappa \pi_a^2}{12a^2\Delta(B)} - \frac{\xi\kappa B \pi_a\pi_B}{a^3\Delta(B)}
- \frac{3\pi_B^2}{2a^4}\left(1 - \frac{3\xi^2 B^2}{2\kappa\Delta(B)}\right)
+ 3a^2 \big(V(B)+\rho\big) \right], \\[6pt]
\dot{\pi}_B &= -\frac{\partial \mathcal{H}_c}{\partial B}
= -N \left[ \frac{\kappa\pi_a^2 \Delta'(B)}{12a\Delta(B)^2}
+ \frac{\xi\kappa \pi_a\pi_B\big(\Delta(B) - B\Delta'(B)\big)}{2a^2\Delta(B)^2}
- \frac{3\xi^2\pi_B^2\big(2B\Delta(B) - B^2\Delta'(B)\big)}{4\kappa a^3\Delta(B)^2}
+ a^3 V'(B) \right].
\end{align}

\subsection{Conservation of Phase Space Volume}
Here, the conservation of the phase-space volume element $dV = da\,dB\,d\pi_a\,d\pi_B$ over time has been carried out. Under Liouville's theorem, a canonical system maintains a constant phase-space volume if the divergence of its phase-space flow field $\mathbf{v} = (\dot{a}, \dot{B}, \dot{\pi}_a, \dot{\pi}_B)$ vanishes:
\begin{equation}
\nabla \cdot \mathbf{v} = \frac{\partial \dot{a}}{\partial a} + \frac{\partial \dot{B}}{\partial B} + \frac{\partial \dot{\pi}_a}{\partial \pi_a} + \frac{\partial \dot{\pi}_B}{\partial \pi_B}.
\label{liouville_div}
\end{equation}
Substituting Hamilton's equations,
\begin{equation}
\nabla \cdot \mathbf{v} = \frac{\partial}{\partial a}\left( \frac{\partial (N\mathcal{H})}{\partial \pi_a} \right) + \frac{\partial}{\partial B}\left( \frac{\partial (N\mathcal{H})}{\partial \pi_B} \right) + \frac{\partial}{\partial \pi_a}\left( -\frac{\partial (N\mathcal{H})}{\partial a} \right) + \frac{\partial}{\partial \pi_B}\left( -\frac{\partial (N\mathcal{H})}{\partial B} \right).
\label{liouville_ham}
\end{equation}
Since $\mathcal{H}$ is continuous and twice continuously differentiable, Clairaut's theorem lets us swap the order of partial derivatives:
\begin{equation}
\nabla \cdot \mathbf{v} = \left[ \frac{\partial^2 (N\mathcal{H})}{\partial a \partial \pi_a} - \frac{\partial^2 (N\mathcal{H})}{\partial \pi_a \partial a} \right] + \left[ \frac{\partial^2 (N\mathcal{H})}{\partial B \partial \pi_B} - \frac{\partial^2 (N\mathcal{H})}{\partial \pi_B \partial B} \right] = 0.
\label{liouville_zero}
\end{equation}
Because the divergence of the flow field vanishes identically, the phase-space volume is strictly conserved, $d(dV)/dt = 0$. This argument depends only on the general Hamiltonian structure $\mathcal{H}_c=N\mathcal{H}$ and is therefore unaffected by the specific form of $\mathcal{H}$; it confirms the mathematical consistency of the canonical framework developed in this work.

\subsection{Geometric Trajectory Restrictions of the Constraint $\mathcal{H} = 0$}
By varying the total canonical action with respect to the lapse function gives
\begin{equation}
\frac{\partial \mathcal{H}_c}{\partial N} = \mathcal{H} = 0.
\label{h_constraint_zero}
\end{equation}
While an unconstrained classical system can move freely through the 4-dimensional phase space $(a, B, \pi_a, \pi_B)$, the condition $\mathcal{H} = 0$ restricts the physical states to a 3-dimensional hypersurface. Furthermore, because the transformation generated by $\mathcal{H}$ corresponds to a gauge reparameterization of cosmic time, points along a given trajectory represent the same physical state under different time coordinates, reducing the true physical phase space to a 2-dimensional reduced manifold.

\section{Stability Analysis and Critical Points}\label{sec4}
This section mainly focuses on testing the stability and finding critical points of the Hamiltonian syatem developed in the previous section. To find the critical (equilibrium) points and analyze their stability, we use the fact that Eqs.~(\ref{adot_final})-(\ref{bdot_final}) together with the momentum equations form a Hamiltonian system governed by $\mathcal{H}_c = N\mathcal{H}$.
An equilibrium point $(\bar a,\bar B,\bar\pi_a,\bar\pi_B)$ occurs when all time derivatives vanish simultaneously:
$$\dot{a} = 0, \quad \dot{B} = 0, \quad \dot{\pi}_a = 0, \quad \dot{\pi}_B = 0.$$
Assuming $N\neq0$ and $a\neq0$: setting $\dot a=0$ in Eq.~(\ref{adot_final}) gives $\pi_a = 3\xi B\pi_B/a$. Substituting into $\dot B=0$ [Eq.~(\ref{bdot_final})] and simplifying, the general solution valid for arbitrary parameters is
$$\bar\pi_a = 0, \qquad \bar\pi_B = 0.$$
Substituting $\pi_a=\pi_B=0$ into the momentum equations of motion then gives
$$-N\left[3a^2(V(B)+\rho)\right]=0 \implies V(\bar B)=-\rho, \qquad -N\left[a^3V'(B)\right]=0 \implies V'(\bar B)=0.$$
The system therefore possesses critical points at
$$(\bar a,\bar B,\bar\pi_a,\bar\pi_B) = (a_0,B_0,0,0),$$
where $B_0$ simultaneously satisfies (1) $V(B_0)=-\rho$ (the potential balances the background density) and (2) $V'(B_0)=0$ (the field sits at an extremum of its potential).

\subsection{Linear Stability}
To determine stability, the linearization is about $(\bar a,\bar B,0,0)$ using the state vector $x=[a,B,\pi_a,\pi_B]^T$ and the Jacobian $J=\partial(\dot a,\dot B,\dot\pi_a,\dot\pi_B)/\partial x$. At the critical point, every term in $\dot a$ and $\dot B$ carries a factor of $\pi_a$ or $\pi_B$, so their derivatives with respect to $a,B$ vanish there; likewise every term in $\dot\pi_a,\dot\pi_B$ that survives at $\pi_a=\pi_B=0,\,V(\bar B)=\rho,\,V'(\bar B)=0$ has vanishing derivative with respect to $\pi_a,\pi_B$. The Jacobian therefore reduces to the block off-diagonal form
$$J = \begin{pmatrix} 0 & 0 & \displaystyle-\frac{N\kappa}{6\bar a\Delta(\bar B)} & \displaystyle\frac{N\xi\bar B}{2\bar a^2\Delta(\bar B)} \\[8pt] 0 & 0 & \displaystyle\frac{N\xi\bar B}{2\bar a^2\Delta(\bar B)} & \displaystyle\frac{N}{\bar a^3}\left(1-\frac{3\xi^2\bar B^2}{2\kappa\Delta(\bar B)}\right) \\[8pt] 0 & 0 & 0 & 0 \\[4pt] 0 & -N\bar a^3 V''(\bar B) & 0 & 0 \end{pmatrix},$$
which have been obtained directly by symbolic differentiation of Hamilton's equations at the critical point rather than by inspection. Solving $\det(J-\lambda I)=0$ gives the characteristic equation
$$\lambda^2\left[\lambda^2 + N^2\bar  V''(\bar B)\,\left(1-\frac{3\xi^2\bar B^2}{2\kappa\Delta(\bar B)}\right)\right]=0,$$
and, after simplification using Eq.~(\ref{delta_def}), the eigenvalues take the closed form
\begin{equation}
\lambda_{1,2}=0,\qquad \lambda_{3,4}=\pm\sqrt{ -N^2\bar  V''(\bar B)\,\left(1-\frac{3\xi^2\bar B^2}{2\kappa\Delta(\bar B)}\right)}.
\label{eigenvalues_corrected}
\end{equation}
The two zero eigenvalues reflect the residual gauge (time-reparameterization) freedom characteristic of constrained Hamiltonian systems in general relativity.
\begin{itemize}
\item \textbf{Stable (oscillatory / center):} if the quantity under the square root is negative, $\lambda_{3,4}=\pm i\omega$ are purely imaginary. This requires
$$ V''(\bar B)\,\left(1-\frac{3\xi^2\bar B^2}{2\kappa\Delta(\bar B)}\right) > 0.$$
\item \textbf{Unstable (saddle):} if the quantity under the square root is positive, then the eigenvalue will have positive real roots and nearby trajectories diverge exponentially.  The condition require for this case is $$ V''(\bar B)\,\left(1-\frac{3\xi^2\bar B^2}{2\kappa\Delta(\bar B)}\right) < 0.$$
\end{itemize}

\section{Dynamic Behavior Near the Lorentz-Violating Vacuum State}\label{sec5}
To study spontaneous Lorentz symmetry breaking within this cosmological framework, we consider a potential $V(B)$ with a Higgs-like symmetry-breaking profile:
\begin{equation}
V(B) = \frac{\lambda}{4}\left( B^\mu B_\mu + b^2 \right)^2 = \frac{\lambda}{4}\left( B^2 - b^2 \right)^2.
\label{higgs_potential}
\end{equation}
The potential minimum defines the vacuum expectation value $\langle B\rangle = b$. As the field settles into this configuration its canonical momentum vanishes, $\pi_B\to0$, and $\Delta(B)$ reduces to the constant
\begin{equation}
\Delta(b) = 1 - \xi b^2 + \frac{3\xi^2 b^2}{2\kappa}.
\label{delta_vacuum}
\end{equation}
Here, emphasizes given on the condition used here is $\pi_B=0$ (vanishing conjugate momentum), evaluated directly in the Hamiltonian constraint~(\ref{hamiltonian_constraint_final}); because the cross term in $\mathcal{H}$ is proportional to $\pi_B$, it drops out automatically regardless of its coefficient, so the results of this section are unaffected by the sign correction of Sec.~III. Substituting $\pi_B=0$ and $\Delta(B)=\Delta(b)$ into Eq.~(\ref{hamiltonian_constraint_final}) gives
\begin{equation}
\mathcal{H}_{\text{vacuum}} = -\frac{\kappa}{12 a \Delta(b)}\pi_a^2 + a^3 \left[ V(b) + \rho \right] = 0.
\label{h_vacuum}
\end{equation}
Setting the time derivatives of $a$ and $\pi_a$ to zero in Hamilton's equations for $\mathcal{H}_{\text{vacuum}}$: from $\dot a = \partial\mathcal{H}_{\text{vacuum}}/\partial\pi_a = -\kappa\pi_a/(6a\Delta(b))$, $\dot a=0$ requires $\bar\pi_a=0$; from $\dot\pi_a=-\partial\mathcal{H}_{\text{vacuum}}/\partial a$, setting $\dot\pi_a=0$ and substituting $\bar\pi_a=0$ gives $-3a^2(V(b)+\rho)=0$, i.e.\ $V(b)=-\rho$ for $a\neq0$. Substituting $\bar\pi_a=0$ back into Eq.~(\ref{h_vacuum}) confirms this is consistent with the constraint. The critical points for the vacuum Hamiltonian~(\ref{h_vacuum}) therefore occur where (1) $\pi_a=0$ (a momentarily static, turning-point configuration) and (2) $V(b)=-\rho$  i.e. the field's potential energy balances the background density.

\subsection{Phase Space Topology and Cosmological Solutions}

To examine the dynamical evolution of the cosmological system governed by the Hamiltonian constraint~(\ref{h_vacuum}), the study of the  phase space in the $(a,\pi_a)$ plane, restricted to $a>0$ has been carried out. Setting $\mathcal{H}_{\text{vacuum}}=E$ for a constant energy hypersurface, the trajectories are implicitly defined by
\begin{equation}
\pi_a^2 = \frac{12\Delta(b)}{\kappa}a^4\big[V(b)+\rho\big] - \frac{12\Delta(b)E}{\kappa}a.
\end{equation}
The qualitative behavior splits into three topological regimes depending on the sign of $\Lambda_{\text{eff}}\equiv V(b)+\rho$.
The physical boundaries of the universe (minimal-size bounce and maximal-size turnaround) occur where the expansion momentarily halts, $\pi_a=0$, giving the real roots of
\begin{equation}
a^3\big[V(b)+\rho\big] = E.
\end{equation}

\subsubsection{Case I: The Under-balanced Universe ($\Lambda_{\text{eff}}<0$)}
When the background density dominates ($V(b)<-\rho$), bounded trajectories exist only for $E<0$. Writing $E=-|E|$,
\begin{equation}
a_{\text{turning}} = \left(\frac{|E|}{|V(b)+\rho|}\right)^{1/3}.
\end{equation}
The system forms closed orbits about the origin of configuration space; the universe avoids the cosmic singularity ($a\to0$), executing non-singular periodic oscillations -- a \textit{cyclic universe}.

\subsubsection{Case II: The Over-balanced Universe ($\Lambda_{\text{eff}}>0$)}
When the potential energy dominates ($V(b)>-\rho$), trajectories are open, unbound power-law branches. On the physical constraint surface ($E=0$),
\begin{equation}
\pi_a = \pm\sqrt{\frac{12\Delta(b)}{\kappa}[V(b)+\rho]}\;a^2.
\end{equation}
\begin{itemize}
\item \textbf{Expanding branch ($\pi_a<0$):} trajectories originate from the Big Bang singularity at $a\to0$ and expand toward $a\to\infty$.
\item \textbf{Contracting branch ($\pi_a>0$):} trajectories evolve from $a\to\infty$ into a Big Crunch singularity at $a=0$.
\end{itemize}

\subsubsection{Case III: The Critical Balance ($\Lambda = 0$) }
When the potential exactly balances the background energy density ($V(b)=-\rho$), the trajectories decouple from the higher-order spatial terms:
\begin{equation}
\pi_a = \pm\sqrt{-\frac{12\Delta(b)E}{\kappa}}\;a^{1/2}.
\end{equation}
This condition forms flat, non-diverging parallel phase traces, marking the critical boundary between the open expanding paths and the closed oscillatory orbits.

\begin{table}[htbp]
\centering
\caption{Classification of cosmological regimes and turning points for Eq.~(\ref{h_vacuum}).}
\label{tab:cosmological_regimes}
\begin{tabular}{l|l|l|l}
\toprule
\textbf{Regime} \hspace{0.5cm}& \textbf{Turning point $(a_{\text{turn}},\pi_a)$}\hspace{0.5cm} & \textbf{Evolution type} \\
\hline
Over-balanced ($\Lambda_{eff}<0$) \hspace{0.5cm}  & \hspace{0.5cm}None (singularity at $a=0$) \hspace{0.5cm}& \hspace{0.5cm}Open Big Bang / Big Crunch \\
Under-balanced ($\Lambda_{eff}>0$)\hspace{0.5cm}  &\hspace{0.5cm} $\left(\left[\frac{|E|}{|\rho+V(b)|}\right]^{1/3},\,0\right)$\hspace{0.5cm} & \hspace{0.5cm}Cyclic / bouncing universe \\
Critical balance ($\Lambda_{eff}=0$)\hspace{0.5cm} & \hspace{0.5cm} None (no turnaround) \hspace{0.5cm}& \hspace{0.5cm}Stable, asymptotically flat \\
\hline
\end{tabular}
\end{table}

\begin{figure}[!h]
\centerline{
  \includegraphics[scale = 0.45]{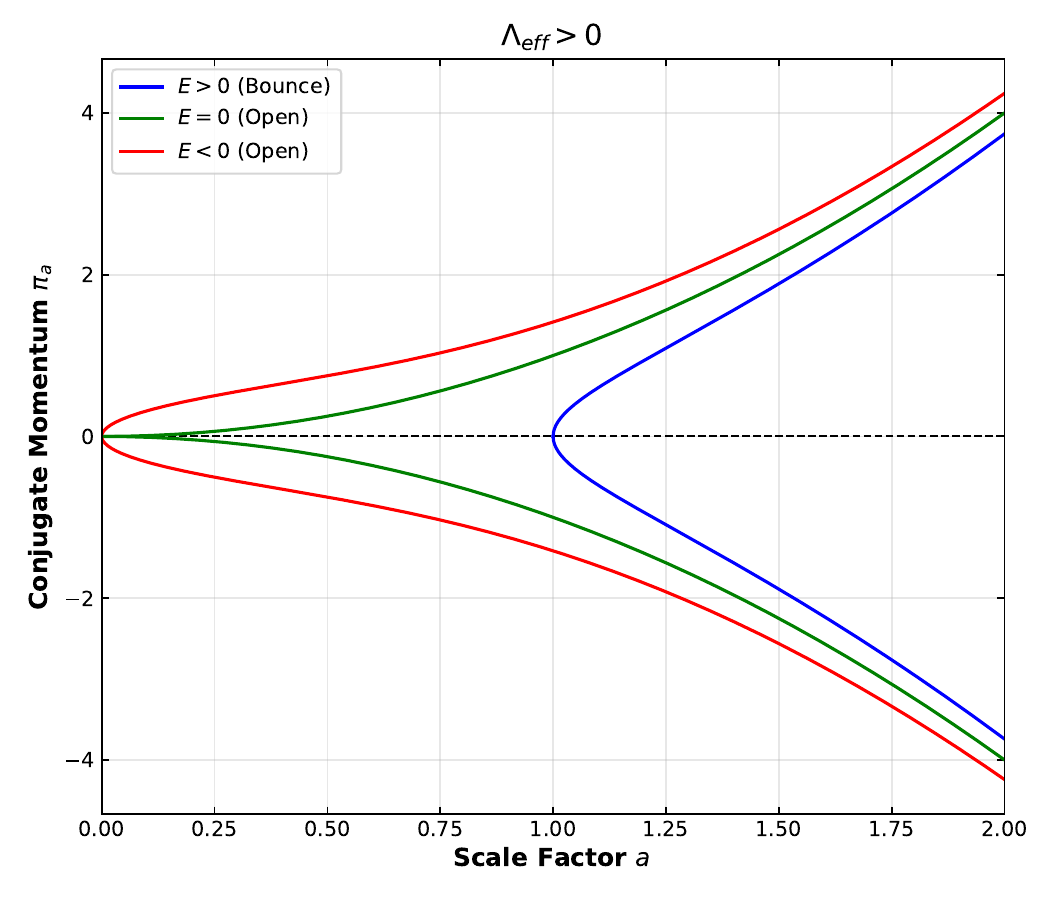}\hspace{0.20cm}
  \includegraphics[scale = 0.45]{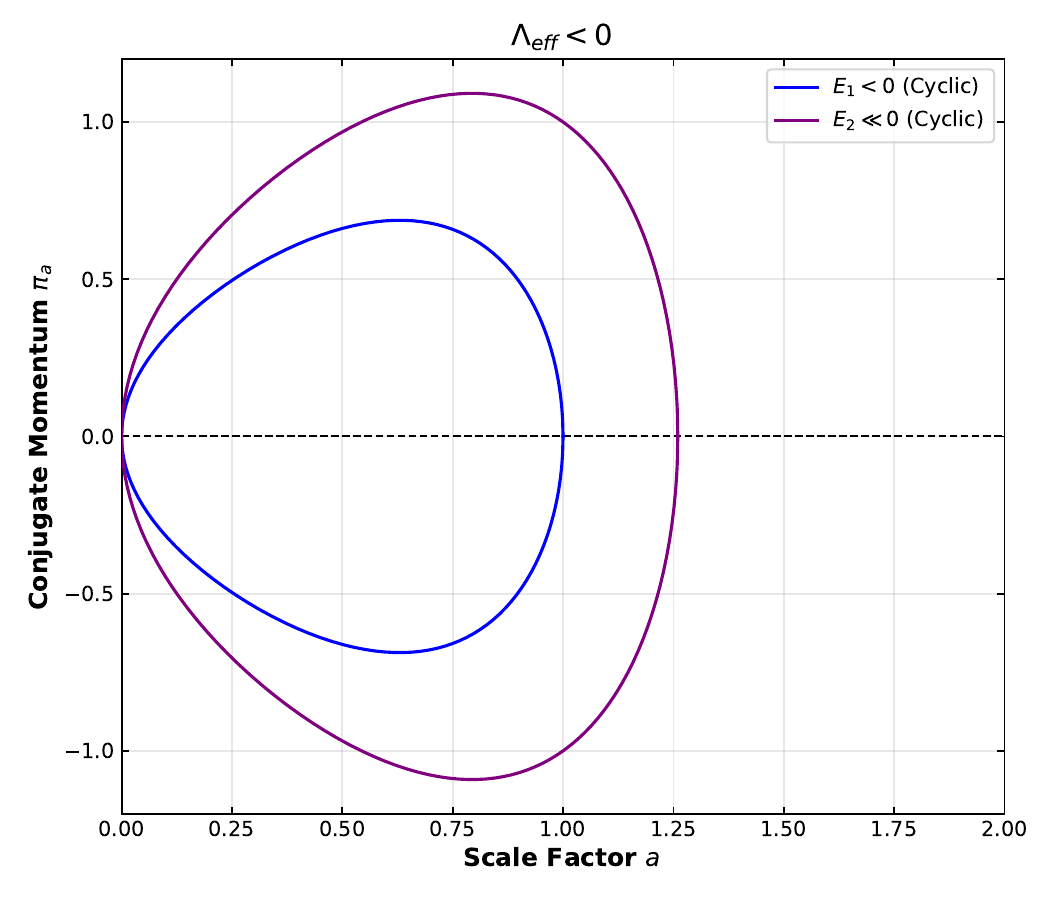}\hspace{0.20cm}
  }
\centerline{
  \includegraphics[scale = 0.45]{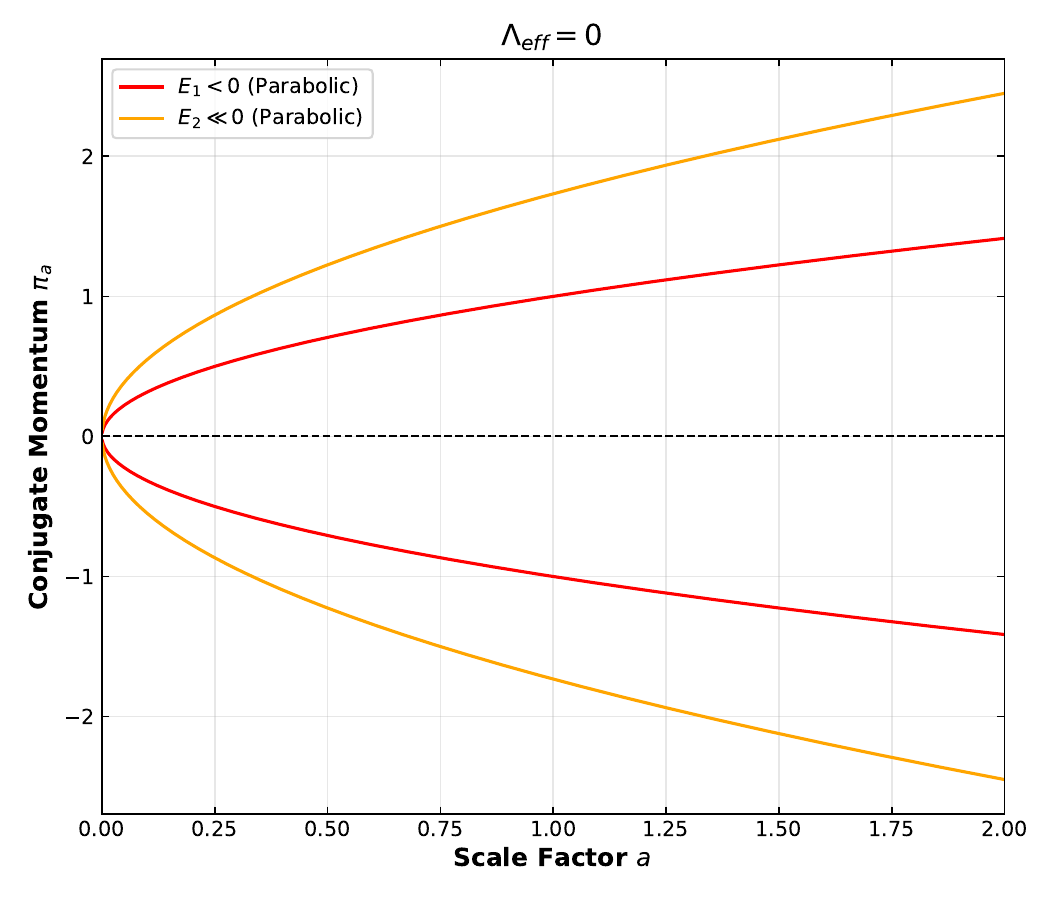}
  }
 \vspace{-0.25cm}
\caption{The schematic graphical behaviour of $(a,\pi_a)$ trajectories on the constraint surface $\mathcal{H}_{\text{vacuum}}=E$, for three different regimes: (a) the over-balanced (open) regime($V(b) < -\rho $), (b) the under-balanced (cyclic/bouncing) regime ($V(b) > -\rho $), and (c) the critical-balance regime ($V(B)(b) = -\rho $).}
\label{fig1}
\vspace{0.5cm} 
\end{figure}

At $V(b)=0$, Eq.~(\ref{h_vacuum}) simplifies to
\begin{equation}
\frac{\kappa}{12a\Delta(b)}\pi_a^2 = a^3\rho,
\label{h_vacuum_final}
\end{equation}
showing that, {\it in this Hamiltonian (momentum-space) description}, settling into the Lorentz-violating vacuum rescales the effective gravitational coupling entering the constraint,
\begin{equation}
\kappa_{\text{eff}} = \frac{\kappa}{\Delta(b)}.
\label{kappa_eff}
\end{equation}
 This statement has been used in Section \ref{sec7}, where it will show that the classical (velocity-space) Friedmann equation instead rescales $\kappa$ by the simpler factor $1/(1-\xi B^2)$, and that the two descriptions coincide only in the idealized limit in which both $\dot B$ and $\pi_B$ vanish -- a limit that, for $\xi\neq0$, strictly requires $\dot a\to0$ as well [see Eq.~(\ref{mom_b})]. Equation~(\ref{kappa_eff}) should therefore be read as the effective coupling appearing in the Hamiltonian constraint at the exact fixed point $\pi_B=0$, rather than as the coefficient multiplying $\rho$ in the classical, time-dependent Friedmann equation away from that fixed point.

\section{The Modified Friedmann and Acceleration Equations}\label{sec6}
To connect the canonical formulation to cosmological observables, we return to the Lagrangian~(\ref{lagrangian_density_final}) and fix the standard cosmic-time gauge $N(t)=1$.

\subsection{The First Friedmann Equation}
The Hamiltonian constraint is equivalent, by construction, to $\partial\mathcal{L}/\partial N=0$ evaluated at $N=1$; we use this direct route, since it avoids re-expressing the momenta in terms of velocities. Varying Eq.~(\ref{lagrangian_density_final}) with respect to $N$ and setting $N=1$ gives, after dividing through by $3a^3/\kappa$ and writing $H\equiv\dot a/a$,
\begin{equation}
H^2\big(1-\xi B^2\big) = \frac{\kappa}{6}\dot B^2 + \xi B\dot B H + \frac{\kappa}{3}\big[V(B)+\rho\big],
\label{friedmann_raw}
\end{equation}
which can be verified independently by symbolic substitution back into the original action. Solving for $H^2$,
\begin{equation}
{H^2 = \frac{\kappa_{eff}}{3}\left[\rho + V(B) + \frac{\dot B^2}{2}\right] + \frac{\xi B\dot B}{1-\xi B^2}\,H}
\label{friedmann_first}
\end{equation}
with the sign of $V(B)$ and $\rho$ inside the bracket fixed by the matter/potential convention $\mathcal{L}_m=-\rho$ used in Eq.~(\ref{lpot_final}) (so that, for $\xi=0$, Eq.~(\ref{friedmann_first}) reduces to the standard result $H^2=\kappa_{eff}~\rho/3$ once the field's own energy density $V(B)+\dot B^2/2$ is moved to the same side as $\rho$). The cross-term $\xi B\dot B H/(1-\xi B^2)$ acts as an interactive drag between the expansion of space and the rolling velocity of the bumblebee field. Because the effective energy density is rescaled by the factor $1/(1-\xi B^2)$, variations in the vector field produce a time-dependent effective gravitational strength, analogous to scalar-tensor or varying-$G$ theories of gravity. We note that the coefficient here is $1-\xi B^2$, and not the Hamiltonian scaling function $\Delta(B)$ of Eq.~(\ref{delta_def}): the latter arises specifically from eliminating $\dot B$ in favor of the canonical momentum $\pi_B$ in the Legendre transformation, whereas Eq.~(\ref{friedmann_first}) is obtained directly in velocity space and is exact for arbitrary $B(t)$.

\subsection{The Second Friedmann Equation}
Differentiating Eq.~(\ref{friedmann_raw}) with respect to cosmic time, using the matter continuity relation $\dot\rho=-3H(\rho+P)$ for a fluid of pressure $P$, and eliminating $\ddot B$ using the field's own equation of motion
\begin{equation}
\ddot B + 3H\dot B + \frac{3\xi B}{\kappa}\frac{\ddot a}{a} + V'(B) = 0.
\end{equation}
 After eliminating the remaining $H^2$ term using Eq.~(\ref{friedmann_raw}) itself, the modified acceleration equation can be obtained as
\begin{equation}
{\frac{\ddot a}{a} = -\frac{\kappa_{eff}}{6}\big(\rho+3P\big) + \frac{\kappa_{eff}}{3}V(B) + \frac{\xi}{2\Delta(B)}(\dot B^2+ B \ddot{B}) - \frac{\kappa_{eff}}{3} \dot{B}^2 + \frac{\xi H B \dot{B}}{\Delta(B)}}
\label{acceleration_equation}
\end{equation}
where $\Delta(B)$ is the function defined in Eq.~(\ref{delta_def}). In standard General Relativity, cosmic acceleration ($\ddot a>0$) requires violation of the strong energy condition ($\rho+3P<0$). In the present model, Eq.~(\ref{acceleration_equation}) shows that, even for ordinary matter satisfying $\rho+3P>0$, the sign and magnitude of $\ddot a/a$ receive competing contributions from the potential $V(B)$, the field's kinetic energy $\dot B^2$, and the cross term containing $\xi B \dot{B} H$, so that accelerated expansion can in principle be driven or assisted by Lorentz-symmetry breaking.

When the field settles into its vacuum state ($B=b$, $\dot B=\ddot B=0$), Eq.~(\ref{acceleration_equation}) simplifies to
\begin{equation}
\frac{\ddot a}{a} = -\frac{\kappa_{eff}}{6}(\rho+3P) + \frac{\kappa_{eff}}{3}V(b),
\label{acceleration_vacuum}
\end{equation}
and the system reduces to a standard FLRW universe governed by the rescaled coupling $\kappa_{\rm eff}=\kappa/\Delta(b)$ of Eq.~(\ref{kappa_eff}), consistent with the Hamiltonian vacuum analysis of Section \ref{sec6}.

In the next  section, the work mainly focuses to the quantum description of this system via canonical quantization of the constraint~(\ref{hamiltonian_constraint_final}).

\section{Canonical Quantization and the Wheeler-DeWitt Equation}\label{sec7}
To transition from classical mechanics to quantum cosmology, we implement canonical quantization by promoting the phase-space coordinates and momenta to operators acting on the wave function of the universe $\Psi(a,B)$:
\begin{equation}
\hat{a} = a, \quad \hat{\pi}_a = -i\hbar \frac{\partial}{\partial a}, \quad \hat{B} = B, \quad \hat{\pi}_B = -i\hbar \frac{\partial}{\partial B},
\label{operators}
\end{equation}
working in natural units $\hbar=1$. Adopting a symmetric operator ordering for hermiticity and promoting $\mathcal{H}=0$ to $\hat{\mathcal{H}}\Psi(a,B)=0$ using the corrected constraint~(\ref{hamiltonian_constraint_final}) gives the Wheeler-DeWitt equation
\begin{equation}
{\left[ \frac{\kappa}{12 a \Delta(B)} \frac{\partial^2}{\partial a^2} - \frac{\xi B}{2 a^2 \Delta(B)} \frac{\partial^2}{\partial a \partial B} - \frac{1}{2a^3} \left( 1 - \frac{3\xi^2 B^2}{2\kappa \Delta(B)} \right) \frac{\partial^2}{\partial B^2} + a^3 \big( V(B) + \rho \big) \right] \Psi(a, B) = 0}
\label{wdw_equation}
\end{equation}
The relative sign between the $\partial^2/\partial a^2$ and $\partial^2/\partial a\partial B$ terms, and the absence of an extra factor of $\kappa$ multiplying $\xi B$ in the mixed derivative coefficient, follow directly from Eq.~(\ref{hamiltonian_constraint_final}) under $\pi_a\to-i\partial_a$, $\pi_B\to-i\partial_B$.

It is to be noted that the WDW equation contains no explicit time derivative $\partial\Psi/\partial t$ as general relativity considers time as a coordinate rather than a background  parameter and also the canonical Hamiltonian gravity satisfies that $\hat H \psi = 0$ \cite{time}. Therefore, time must be recovered relationally, by considering changes in the scale factor $a$ against the evolution of the bumblebee field $B$. Near the classical Big Bang singularity ($a\to0$), the differential coefficients of Eq.~(\ref{wdw_equation}) diverge, and whether $\Psi(a,B)$ vanishes there (the DeWitt boundary condition) determines whether quantum gravity regularizes the classical singularity.

\subsection{Approximate Solutions of the Wheeler-DeWitt Equation}\label{sec8}
Equation~(\ref{wdw_equation}) is a second-order partial differential equation with variable coefficients and a genuine mixed-derivative term, and no closed-form solution valid for all $(a,B)$ is available for any form of generic potential $V(B)$. In this work, the analysis has been carried  out for two approximations: the deep-quantum regime near the classical singularity, and the small-oscillation regime near the Lorentz-violating vacuum.

Here, expanding around the VEV, $B=b+\chi$ with $\chi$ as a small fluctuation, the Higgs-like potential~(\ref{higgs_potential}) reduces to
\begin{equation}
V(B) = \frac{\lambda}{4}\big(2b\chi+\chi^2\big)^2 \approx \lambda b^2\chi^2 + \mathcal{O}(\chi^3),
\end{equation}
and, evaluating the coefficients of Eq.~(\ref{wdw_equation}) at $B\approx b$ so that $\Delta(B)\approx\Delta_0\equiv\Delta(b)$, we define the constants
\begin{equation}
\gamma=\frac{\kappa}{12\Delta_0},\qquad \eta=\frac{\xi b}{2\Delta_0},\qquad \zeta=\frac{1}{2}\left(1-\frac{3\xi^2b^2}{2\kappa\Delta_0}\right).
\end{equation}
Multiplying Eq.~(\ref{wdw_equation}) by $a^3$ and introducing the logarithmic scale factor $\alpha=\ln a$ (so that $a\,\partial_a = \partial_\alpha$ and $a^2\partial_a^2=\partial_\alpha^2-\partial_\alpha$), the equation becomes
\begin{equation}
\left[\gamma\left(\frac{\partial^2}{\partial\alpha^2}-\frac{\partial}{\partial\alpha}\right) - \eta\frac{\partial^2}{\partial\alpha\,\partial\chi} - \zeta\frac{\partial^2}{\partial\chi^2} + e^{6\alpha}\big(\lambda b^2\chi^2+\rho\big)\right]\Psi(\alpha,\chi)=0.
\label{wdw_alpha}
\end{equation}

\subsubsection{Deep-quantum regime ($a\to0$)}
As $\alpha\to-\infty$ the exponential potential term vanishes; retaining the dominant second-derivative behavior and dropping the subleading first-order drift term $\gamma\partial_\alpha$, Eq.~(\ref{wdw_alpha}) reduces to the homogeneous, constant-coefficient hyperbolic equation
\begin{equation}
\gamma\frac{\partial^2\Psi}{\partial\alpha^2} - \eta\frac{\partial^2\Psi}{\partial\alpha\,\partial\chi} - \zeta\frac{\partial^2\Psi}{\partial\chi^2} = 0.
\end{equation}
Substituting a plane-wave ansatz $\Psi\propto e^{i(k_\alpha\alpha+k_\chi\chi)}$ gives the characteristic equation
\begin{equation}
\gamma k_\alpha^2 - \eta k_\alpha k_\chi - \zeta k_\chi^2 = 0 \quad\Longrightarrow\quad k_\alpha = \frac{\eta\pm\sqrt{\eta^2+4\gamma\zeta}}{2\gamma}\,k_\chi \equiv m_\pm\,k_\chi.
\label{char_roots}
\end{equation}
Because $\gamma,\zeta>0$ for $\xi^2b^2<2\kappa/3$, the discriminant $\eta^2+4\gamma\zeta$ is manifestly positive and $m_\pm$ are real for all coupling strengths in this range -- a direct consequence of the corrected (positive) sign of the mixed term, and a qualitative difference from what an incorrectly-signed cross term would imply. The general solution near the boundary is a superposition of the two characteristic wave families,
\begin{equation}
\Psi(\alpha,\chi) = \int_{-\infty}^{\infty}\Big[C_1(k)\,e^{ik(\chi+m_+\alpha)} + C_2(k)\,e^{ik(\chi+m_-\alpha)}\Big]\,dk.
\end{equation}
The DeWitt boundary condition requires $\Psi\to0$ as $\alpha\to-\infty$; since $m_\pm$ are real here, this cannot be enforced by simple exponential damping of individual plane waves, and instead requires the superposition $C_1,C_2$ to interfere destructively at the boundary -- e.g., through an odd (sine-type) combination $\Psi\propto\int_0^\infty A(k)\sin\!\big[k(m_+-m_-)\alpha\big]e^{ik\chi}\,dk$, which vanishes identically at $\alpha\to-\infty$ term wise only in a distributional/wave-packet sense. We record this as the natural candidate boundary behavior rather than as a rigorously constructed normalizable state; a fully quantitative treatment (e.g.\ via a wave-packet construction with a specified spectral weight $A(k)$)  can be derived, which has not carried out in this work.

\subsubsection{Near-vacuum harmonic branch}
Away from the immediate boundary, separating $\Psi(\alpha,\chi)=X(\alpha)Y(\chi)$ in Eq.~(\ref{wdw_alpha}) and neglecting the mixed term (valid when $\xi b\ll1$, so that $\eta\ll\sqrt{\gamma\zeta}$) gives a decoupled pair of ordinary differential equations. The $\chi$-equation is that of a quantum harmonic oscillator with effective mass parameter $M^2=\lambda b^2/\zeta$,
\begin{equation}
\left[-\zeta\frac{d^2}{d\chi^2} + e^{6\alpha}\lambda b^2\chi^2\right]Y(\chi) = E_n Y(\chi)
\end{equation}
Here, the standard solution of this quantum harmonic oscillator will be $Y_n(\chi) \propto N_n\,H_n(\gamma_0 e^{{3\alpha}/{2}}\chi)\,e^{-(\gamma_0^2e^{3\alpha}\chi^2)/2},$ with $\gamma_0=\left(\frac{\lambda b^2}{\zeta}\right)^{1/4},\;\; E_n = \sqrt{\zeta\lambda b^2} e^{3\alpha}\left(2n+1\right),$
and $H_n$ is the Hermite polynomials. Now expanding $\Psi=\sum_nX_n(\alpha)Y_n(\chi;\alpha)$ and neglecting the mixed term ($\eta\ll\sqrt{\gamma\zeta}$, i.e.\ $\xi b\ll1$) and the non-adiabatic couplings
$\partial_\alpha Y_n$, The remaining equation will be an explicit $\alpha$ equation of the form,
\begin{equation}
\left[\gamma\left(\frac{d^2}{d\alpha^2}-\frac{d}{d\alpha}\right) + \rho\,e^{6\alpha}\right]X(\alpha) = E_n X(\alpha),
\end{equation}
is, after the substitution $a=e^\alpha$, an exponential-potential (Bessel-type) ordinary differential equation of the same family that appears throughout minisuperspace quantum cosmology for a constant-density source~\cite{Bertolami_2005}; its solutions can be written in terms of Bessel functions of imaginary order, $X(a)\propto a^{1/2}\big[C_1 K_{i\nu}(z)+C_2 I_{i\nu}(z)\big]$ with $z\propto a^3$ and $\nu$ set by $E_n$. Here, $ I_{i\nu}$ and $K_{i\nu}$ are the Bessel function of first and second kind with $\nu = \sqrt{-\frac{1}{9} (\frac{E_n}{\gamma}+ \frac{1}{4})}$.
Taken together, the destructively-interfering boundary state near $a\to0$ and the discretized, harmonically-confined behavior of the field fluctuation $\chi$ around the vacuum are consistent with a quantum resolution of the classical singularity together with a stable, quantized bumblebee field. 

\subsection{Semiclassical WKB Approximation and Singularity Resolution}
To trace the quantum-to-classical transition and confirm the structural consistency of the exact analytical solutions,  implementation of  semiclassical WKB matching analysis within the high-energy ultraviolet domain is a good consideration. In terms of the cosmological scale factor $a$, the effective Wheeler-DeWitt equation under the constant-density background asymptotically reduces to the following differential form as $a \to 0$:
\begin{equation}
    \frac{d^2 X(a)}{da^2} + \left( \frac{9\nu^2 - \frac{1}{4}}{a^2} \right) X(a) = 0.
\end{equation}
By executing the standard semiclassical ansatz $X(a) = \exp[i S(a)/\hbar]$ and isolating the leading phase-integral configuration inside the classically allowed regime, the corresponding WKB momentum field matches $p(a) \approx \pm 3\nu/a$. The total integrated phase contribution yields,
\begin{equation}
    S(a) \approx \pm \int p(a) \, da = \pm 3\nu \ln(a).
\end{equation}
Accounting for the essential probability conservation amplitude factor, $1/\sqrt{p(a)} \propto a^{1/2}$, the general oscillatory semiclassical wave matching the initial boundary conditions settles into the closed real form:
\begin{equation}
    X_{\text{WKB}}(a) = \mathcal{A} \, a^{1/2} \cos\left(3\nu \ln(a) - \phi_0\right),
\end{equation}
where $\mathcal{A} \approx \sqrt{\frac{\pi}{\nu \sinh(\pi\nu)}}$ represents the asymptotic normalization constant mapping directly onto the exact modified Bessel framework, and $\phi_0$ is a dynamic phase shift setting the phase alignment boundary.

The structural transition between dynamic quantum fluctuations and late-time stable classical regimes is strictly governed by the classical turning point $a_{\text{tp}}$, explicitly defined by the parameter scaling:
\begin{equation}
    a_{\text{tp}} = \nu^{1/3}.
\end{equation}
For values deeply embedded inside the classically allowed zone ($a < a_{\text{tp}}$), the wave function undergoes infinitely rapid phase variations proportional to $\ln(a)$ alongside a power-law amplitude damping near the Big Bang boundary ($a \to 0$), successfully demonstrating the singularity resolution. Conversely, across the barrier boundary ($a > a_{\text{tp}}$), the oscillatory behavior terminates sharply, shifting continuously into the classically forbidden Euclidean tunneling domain dominated by exponential WKB tracking tails.
\begin{figure}[!h]
\centerline{
  \includegraphics[scale = 0.55]{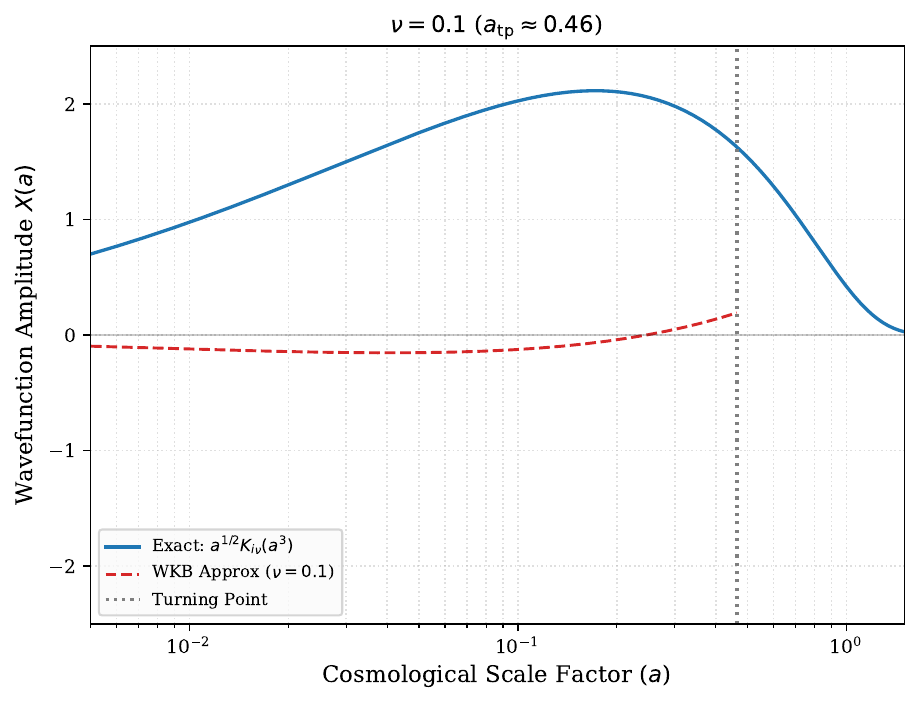 }\hspace{0.20cm}
  \includegraphics[scale = 0.55]{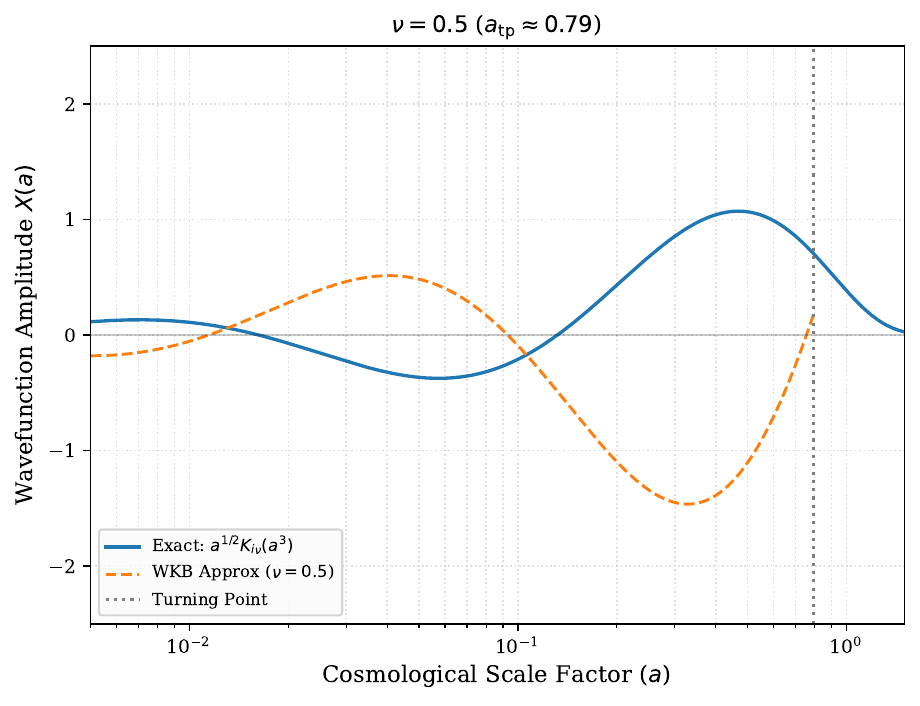 }\hspace{0.20cm}
  }
\centerline{
  \includegraphics[scale = 0.55]{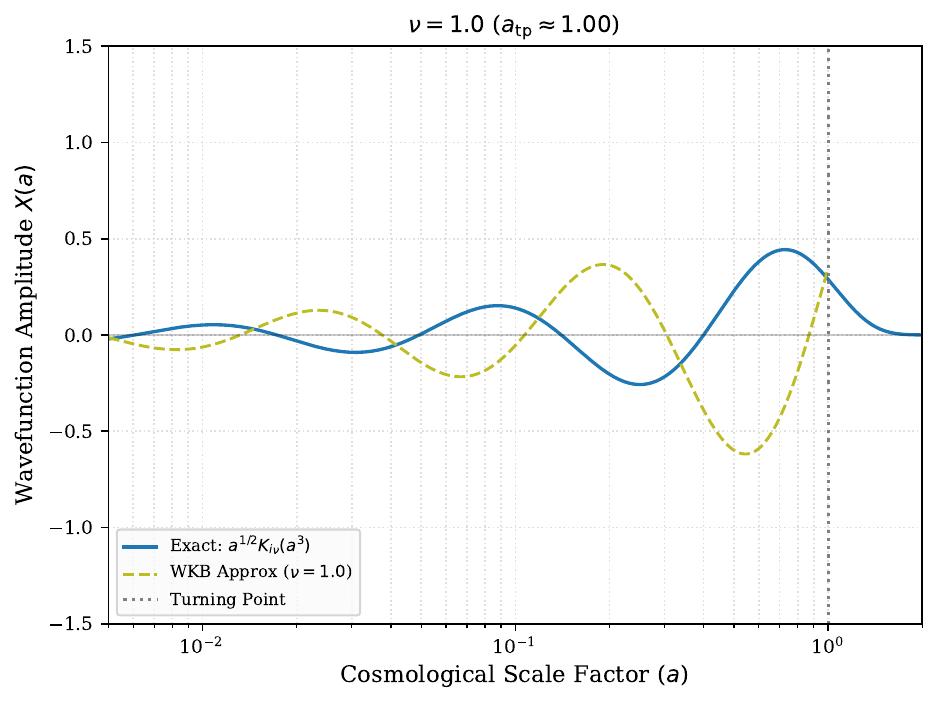 }\hspace{0.20cm}
  \includegraphics[scale = 0.55]{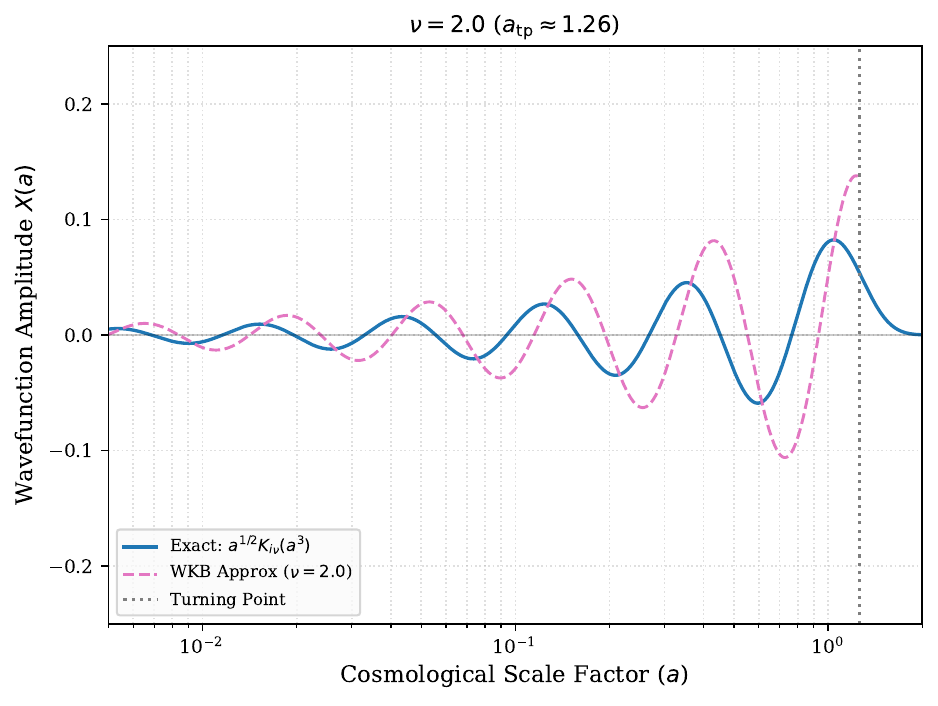 }\hspace{0.20cm}
  }
 \vspace{-0.25cm}
\caption{The Semiclassical transition signatures of the cosmological wavefunction \(X(a)\) evaluated at parameter \(\nu = 0.1, 0.5, 1.0, 2.0\) for $\phi_0 \neq 0$. The pink vertical dotted line clarifies the classical turning point boundary where the active kinetic oscillatory domain terminates.}
\label{fig2}
\vspace{0.5cm} 
\end{figure}

Figure \ref{fig2} illustrates thee exact Bessel-type configuration and its semiclassical WKB counterpart safely below the turning point barrier for $\phi_0 \neq 0$. From the figure, it has been noticed that as $\nu$ increases to higher value, the semiclassical WKB approximation tends to follow the Bessel function exact solution. It has been also observed that, with the increase of $\nu$, the classical turning point also inverses.


\section{Conclusion}\label{sec8}

The work mainly focus on to construct a canonical Hamiltonian system to study the isotropic Bumblebee mini-superspace cosmological system in presence of temporal bumblebee field and further extended it to quantum mechanical analysis. The entire work can be subdivided in to two parts. The first part starting from the covariant action, we have derived the reduced mini-superspace Lagrangian density thorough step by step in section \ref{sec2} and then  carried out the derivation to construct the classical Hamiltonian framework in section \ref{sec3}. To verify the internal consistency of this framework, the conservation of phase-space volume via Louisville's theorem has been tested. The Section \ref{sec4}  contains the study of  the critical points of the Hamiltonian system through linear stability analysis and draw some conclusions regarding the nature of the critical points while analysing the eigen values of the Jacobian. The analysis  shows that while two out of four eigen values are zero and the remaining two are imaginary in nature. With the observation of these characteristics,different conditions for which they behave like either stable (oscillatory/center) or unstable (saddle) has been studied too. The work further extended to the analysis for a Higgs-like potential near Lorentz symmetry breaking state in Section \ref{sec5}. Here, we have derived the Hamiltonian by taking the constant expectation value for the bumblebee field (i.e., $<B> = b$ ) and studied the phase space topology. We analysed three cases which includes The Under-balanced, Over-balanced and Critical balance condition of the Universe. During the analysis we have also pointed that  the effective gravitational coupling is rescaled from $\kappa \to \kappa_{eff}$, in the Hamiltonian (momentum-space) and Lagrangian (velocity-space) descriptions, for the exact fixed point $\dot B=\pi_B=0$. Where $\kappa_{eff} = \frac{\kappa}{1-\xi B^2}$. In Section \ref{sec6}, the modified for of the first and second Friedmann Equations have been derived for the rescaled effective gravitational coupling near Lorentz symmetry breaking state. 

The second part mainly consider quantum cosmological scenario, which has been extensively studied in Section \ref{sec7}. The section starting with defining important phase space coordinates and momenta as quantum mechanical operators and derive the Wheeler-DeWitt (WDW) equation. The approximate solution of the WDW equation has been discussed for the Higgs-like potential of  Section \ref{sec5} for three Deep quantum regime and the near-vacuum harmonic branch.  For the deep quantum regime, the solution is sine type combination of the form $\Psi\propto\int_0^\infty A(k)\sin\!\big[k(m_+-m_-)\alpha\big]e^{ik\chi}\,dk$ and for near-harmonic branch the solution can be written in the form of Bessel function of imaginary order, i.e. $X(a)\propto a^{1/2}\big[C_1 K_{i\nu}(z)+C_2 I_{i\nu}(z)\big]$. Furthermore, the work  carried out the semiclassical WKB approximation and derive the reduced form of WDW equation and the solution obtained has the form  $X_{\text{WKB}}(a) = \mathcal{A} \, a^{1/2} \cos\left(3\nu \ln(a) - \phi_0\right)$, with $\mathcal{A} \approx \sqrt{\frac{\pi}{\nu \sinh(\pi\nu)}}$. Apart from thia, the work also compared the exact solution of the wavefunction for near -vacuum branch and Bessel function form of WKB solution for $\phi_o \neq 0$ and we observed that higher value of $\nu$ the Bessel function  solutions tends to follow the exact solution as shown in Fig \ref{fig2}. In addition to it, there has been observed that the classical turning point also increase with the increasing values of $\nu$.

The isotropic cosmological system in presence of temporal bumblebee field may provides better understanding as well as new insight to the  evolution and dynamics of the Universe from its early stage. With more observational data  form  early cosmos will definitively help  the researchers to test the existencence of field and potential considered in this work and its effect in cosmological system.

\end{document}